\documentstyle[prd,eqsecnum,aps,12pt]{revtex}
\begin{document}
\draft
\title{Nonlinear couplings and tree amplitudes in \\
      gauge theories}

\author{F. T. Brandt and J. Frenkel}
\address{Instituto de F\'\i sica, Universidade de S\~ao Paulo,
S\~ao Paulo, 05389-970 SP, Brasil}

\date{\today}

\maketitle

\vskip 1.0cm

\begin{abstract}
Following a remark advanced by Feynman,
we study the connection between the form
of the nonlinear vertices involving
gauge particles and the Abelian gauge invariance of physical tree amplitudes.
We show that this requirement, together with some natural assumptions,
fixes uniquely the structure of the Yang-Mills theory.
However, the constraints imposed by the above property are not
sufficient to single out the gauge theory of gravitation.
\end{abstract}

\pacs{}

\section{Introduction}

In the Yang-Mills theory, the source of the Yang-Mills fields is the
conserved color current. Since these fields carry color, these will
self-interact leading to a non-Abelian gauge theory \cite{yangmills}.
Similarly, the source of the gravitational fields is the energy-momentum
tensor, a quantity which is locally conserved. These fields carry energy
and momentum and hence must couple to themselves. The non-Abelian gauge
theory of gravitation, which is invariant under local gauge
transformations, is identical to Einstein's theory \cite{weinberg}.
There has been much fundamental work on basic aspects of the non-Abelian
gauge theories
\cite{feynman,dewitt,faddeev,mandelstam,fradkin,thooft,thooftveltman}.
Feynman \cite{feynman} has shown that in these theories, the tree amplitudes
involving free external gauge fields must be invariant under Abelian
gauge transformations of the external fields. He remarked that this
property may be used
in order to investigate, in an alternative way, the structure of the
nonlinear graviton interactions.

The purpose of this work is to study the question whether the above
property of physical tree amplitudes is sufficient to determine completely
the form of the nonlinear interactions between the gauge particles. We consider
this problem in section \ref{sec2}, first in the simpler context of the
Yang-Mills theory. We assume that the nonlinear interactions between the
gluons are local and involve only dimensionless coupling constants. We find
that in this case the answer to the above question is affirmative, basically
due to the absence of gluon vertices of higher degree than four. In section
\ref{sec3}, we work out the corresponding expressions for gravity, whose
algebraic complexity is much greater. We assume that the interactions
between the gravitons are local and involve only two derivatives of these
fields. This allows for the presence of graviton self-couplings to all
orders. In the gravity case, it is always possible to make a local
redefinition of the basic fields, such that the physical amplitudes will
be the same \cite{thooftveltman}. We argue that, even accounting for this
possibility, the Abelian gauge invariance of the tree amplitudes does not
yield enough constraints to fix the form of the nonlinear
graviton couplings.

We report for simplicity only the results for pure gauge theories, since
the problem we study is basically connected with the self-interaction of
gauge particles. We have verified that the introduction of matter fields
adds only a further algebraic complication, without modifying the above
conclusions. Finally, we mention that other interesting aspects of tree
amplitudes in gauge theories have been discussed recently in the literature
\cite{choi,gould,bern}.

\section{The Yang-Mills theory}\label{sec2}

We start with the Yang-Mills case, characterized by a gauge field
$A^a_\alpha$, where $a$ denotes the color index and $\alpha$ is a
Lorentz index.
The quadratic part of the Yang-Mills Lagrangian
\begin{equation}\label{eq2.1}
{\cal L}^2_{YM}\left(A\right)=
\frac{1}{4}\left(\partial_\beta A^a_\alpha -\partial_\alpha A^a_\beta \right)
           \left(\partial_\beta A^a_\alpha -\partial_\alpha A^a_\beta \right),
\end{equation}
is invariant under the Abelian gauge transformation
\begin{equation}\label{eq2.2}
A^a_\alpha\rightarrow A^a_\alpha + \partial_\alpha \omega^a.
\end{equation}
This leads in momentum space to the free equation of motion
\begin{equation}\label{eq2.3}
\left(\eta_{\alpha\beta}k^2-k_\alpha k_\beta\right)A_\beta^a\left(k\right)=0,
\end{equation}
which is invariant under the gauge transformation
\begin{equation}\label{eq2.4}
\delta A^a_\alpha\left(k\right)=\omega^a k_\alpha.
\end{equation}

We now consider the interactions between the gluons, which we assume to be
local and characterized by dimensionless coupling constants. This natural
assumption allows for vertices involving 3 gluons with one derivative term and
4 gluons with no derivatives, but precludes the presence of higher order
gluon self-couplings. In this case, using Bose symmetry and Lorentz
invariance and disregarding total derivatives terms, we can write the
interaction Lagrangian as follows
\begin{eqnarray}\label{eq2.5}
{\cal L}^I_{YM}\left(A\right)&=&
\left(g\,f_{abc} + e_0\,d_{abc}\right)\left(\partial_\nu A^a_\mu\right)
A^b_\mu A^c_\nu +\nonumber \\
& & \left(l_0\,f_{abe} f_{cde} + l_1\,d_{abe} d_{cde} +
    l_2\,\delta_{ab} \delta_{cd}\right) A^a_\mu  A^b_\nu A^c_\mu A^d_\nu +
\nonumber \\
& & \left(l_3\,d_{abe} d_{cde} + l_4\,\delta_{ab} \delta_{cd}\right)
A^a_\mu  A^b_\mu A^c_\nu A^d_\nu.
\end{eqnarray}
Here $f_{abc}$ denote the antisymmetric color structure constants of the gauge
group SU(N) and $d_{abc}$ are the symmetric color factors. The coupling
constant $g$ sets the scale of the gluon interactions and $e_0$, $l_i$ are
dimensionless couplings which must be determined.

\input prepictex
\input pictex
\input postpictex
\begin{figure}
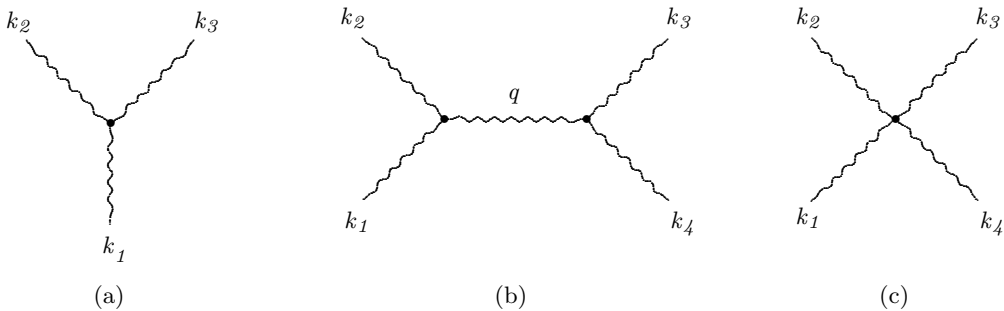

\font\thinlinefont=cmr5
\begingroup\makeatletter\ifx\SetFigFont\undefined
\def\x#1#2#3#4#5#6#7\relax{\def\x{#1#2#3#4#5#6}}%
\expandafter\x\fmtname xxxxxx\relax \def\y{splain}%
\ifx\x\y   
\gdef\SetFigFont#1#2#3{%
  \ifnum #1<17\tiny\else \ifnum #1<20\small\else
  \ifnum #1<24\normalsize\else \ifnum #1<29\large\else
  \ifnum #1<34\Large\else \ifnum #1<41\LARGE\else
     \huge\fi\fi\fi\fi\fi\fi
  \csname #3\endcsname}%
\else
\gdef\SetFigFont#1#2#3{\begingroup
  \count@#1\relax \ifnum 25<\count@\count@25\fi
  \def\x{\endgroup\@setsize\SetFigFont{#2pt}}%
  \expandafter\x
    \csname \romannumeral\the\count@ pt\expandafter\endcsname
    \csname @\romannumeral\the\count@ pt\endcsname
  \csname #3\endcsname}%
\fi
\fi\endgroup
\mbox{\beginpicture
\setcoordinatesystem units < 0.350cm, 0.350cm>
\unitlength= 0.350cm
\linethickness=1pt
\setplotsymbol ({\makebox(0,0)[l]{\tencirc\symbol{'160}}})
\setshadesymbol ({\thinlinefont .})
\setlinear
%
%
\linethickness= 0.500pt
\setplotsymbol ({\thinlinefont .})
\put{\makebox(0,0)[l]{\circle*{ 0.318}}}
at 17.748 15.272
%
\linethickness= 0.500pt
\setplotsymbol ({\thinlinefont .})
\put{\makebox(0,0)[l]{\circle*{ 0.318}}}
at 23.146 15.272
%
\linethickness= 0.500pt
\setplotsymbol ({\thinlinefont .})
\put{\makebox(0,0)[l]{\circle*{ 0.318}}}
at 34.893 15.272
%
\linethickness= 0.500pt
\setplotsymbol ({\thinlinefont .})
\put{\makebox(0,0)[l]{\circle*{ 0.318}}}
at  5.080 15.145
\linethickness= 0.500pt
\setplotsymbol ({\thinlinefont .})
%
%
%
\plot	17.748 15.272 17.907 15.272
 	17.986 15.277
	18.066 15.292
	18.145 15.316
	18.224 15.351
	18.304 15.376
	18.383 15.371
	18.463 15.336
	18.542 15.272
	18.621 15.212
	18.701 15.192
	18.780 15.212
	18.860 15.272
	18.939 15.331
	19.018 15.351
	19.098 15.331
	19.177 15.272
	19.256 15.212
	19.336 15.192
	19.415 15.212
	19.494 15.272
	19.574 15.331
	19.653 15.351
	19.733 15.331
	19.812 15.272
	19.891 15.212
	19.971 15.192
	20.050 15.212
	20.130 15.272
	20.209 15.331
	20.288 15.351
	20.368 15.331
	20.447 15.272
	20.526 15.212
	20.606 15.192
	20.685 15.212
	20.765 15.272
	20.844 15.331
	20.923 15.351
	21.003 15.331
	21.082 15.272
	21.161 15.212
	21.241 15.192
	21.320 15.212
	21.400 15.272
	21.479 15.331
	21.558 15.351
	21.638 15.331
	21.717 15.272
	21.796 15.212
	21.876 15.192
	21.955 15.212
	22.035 15.272
	22.114 15.331
	22.193 15.351
	22.273 15.331
	22.352 15.272
	22.431 15.207
	22.511 15.173
	22.590 15.168
	22.670 15.192
	22.749 15.227
	22.828 15.252
	22.908 15.267
	22.987 15.272
	 /
\plot 22.987 15.272 23.146 15.272 /
\linethickness= 0.500pt
\setplotsymbol ({\thinlinefont .})
%
%
%
\plot	17.748 15.272 17.605 15.145
 	17.540 15.081
	17.486 15.018
	17.415 14.891
	17.328 14.796
	17.254 14.772
	17.161 14.764
	17.071 14.751
	17.006 14.712
	16.967 14.648
	16.954 14.557
	16.943 14.467
	16.907 14.403
	16.847 14.364
	16.764 14.351
	16.681 14.337
	16.621 14.295
	16.585 14.226
	16.574 14.129
	16.558 14.032
	16.510 13.962
	16.431 13.920
	16.320 13.906
	16.209 13.892
	16.133 13.847
	16.090 13.773
	16.081 13.668
	16.074 13.564
	16.038 13.490
	15.971 13.445
	15.875 13.430
	15.778 13.416
	15.708 13.375
	15.667 13.305
	15.653 13.208
	15.639 13.111
	15.597 13.041
	15.528 13.000
	15.430 12.986
	15.333 12.972
	15.264 12.930
	15.222 12.861
	15.208 12.764
	15.198 12.663
	15.169 12.585
	15.050 12.494
	14.973 12.466
	14.903 12.430
	14.780 12.335
	 /
\plot 14.780 12.335 14.669 12.224 /
\linethickness= 0.500pt
\setplotsymbol ({\thinlinefont .})
%
%
%
\plot	26.257 12.192 26.114 12.319
 	26.043 12.377
	25.971 12.422
	25.900 12.456
	25.829 12.478
	25.721 12.553
	25.695 12.624
	25.686 12.716
	25.673 12.806
	25.634 12.871
	25.570 12.909
	25.479 12.922
	25.388 12.935
	25.321 12.974
	25.277 13.038
	25.257 13.129
	25.237 13.220
	25.194 13.287
	25.126 13.331
	25.035 13.351
	24.944 13.373
	24.876 13.422
	24.832 13.500
	24.813 13.605
	24.792 13.709
	24.745 13.783
	24.673 13.828
	24.575 13.843
	24.476 13.856
	24.404 13.895
	24.357 13.959
	24.336 14.049
	24.317 14.141
	24.273 14.208
	24.205 14.252
	24.114 14.272
	24.023 14.291
	23.955 14.335
	23.912 14.403
	23.892 14.494
	23.872 14.585
	23.828 14.653
	23.761 14.696
	23.670 14.716
	23.576 14.733
	23.503 14.768
	23.416 14.891
	23.388 14.967
	23.352 15.038
	23.257 15.161
	 /
\plot 23.257 15.161 23.146 15.272 /
\linethickness= 0.500pt
\setplotsymbol ({\thinlinefont .})
%
%
%
\plot	37.973 18.320 37.830 18.193
 	37.765 18.129
	37.711 18.066
	37.640 17.939
	37.552 17.844
	37.479 17.820
	37.386 17.812
	37.295 17.799
	37.231 17.760
	37.192 17.696
	37.179 17.605
	37.167 17.515
	37.132 17.451
	37.072 17.412
	36.989 17.399
	36.905 17.385
	36.846 17.343
	36.810 17.274
	36.798 17.177
	36.782 17.080
	36.735 17.010
	36.655 16.968
	36.544 16.954
	36.434 16.940
	36.358 16.895
	36.315 16.821
	36.306 16.716
	36.299 16.612
	36.262 16.538
	36.196 16.493
	36.100 16.478
	36.003 16.464
	35.933 16.423
	35.891 16.353
	35.877 16.256
	35.864 16.159
	35.822 16.089
	35.752 16.048
	35.655 16.034
	35.558 16.020
	35.489 15.978
	35.447 15.909
	35.433 15.811
	35.423 15.711
	35.393 15.633
	35.274 15.542
	35.198 15.514
	35.127 15.478
	35.004 15.383
	 /
\plot 35.004 15.383 34.893 15.272 /
\linethickness= 0.500pt
\setplotsymbol ({\thinlinefont .})
%
%
%
\plot	34.893 15.272 35.036 15.145
 	35.108 15.087
	35.179 15.042
	35.250 15.008
	35.322 14.986
	35.429 14.911
	35.456 14.840
	35.465 14.748
	35.478 14.658
	35.516 14.593
	35.581 14.554
	35.671 14.541
	35.762 14.529
	35.830 14.490
	35.874 14.425
	35.893 14.335
	35.913 14.244
	35.957 14.176
	36.024 14.133
	36.116 14.113
	36.207 14.091
	36.274 14.041
	36.318 13.964
	36.338 13.859
	36.359 13.755
	36.405 13.680
	36.478 13.636
	36.576 13.621
	36.674 13.608
	36.747 13.569
	36.793 13.505
	36.814 13.414
	36.834 13.323
	36.878 13.256
	36.945 13.212
	37.036 13.192
	37.128 13.172
	37.195 13.129
	37.239 13.061
	37.259 12.970
	37.278 12.879
	37.322 12.811
	37.390 12.767
	37.481 12.748
	37.574 12.731
	37.648 12.696
	37.735 12.573
	37.763 12.497
	37.798 12.426
	37.894 12.303
	 /
\plot 37.894 12.303 38.005 12.192 /
\linethickness= 0.500pt
\setplotsymbol ({\thinlinefont .})
%
%
%
\plot	31.686 12.160 31.829 12.287
 	31.895 12.351
	31.948 12.414
	32.020 12.541
	32.107 12.636
	32.181 12.660
	32.274 12.668
	32.364 12.681
	32.429 12.720
	32.467 12.784
	32.480 12.875
	32.492 12.965
	32.528 13.029
	32.587 13.068
	32.671 13.081
	32.754 13.095
	32.814 13.137
	32.849 13.206
	32.861 13.303
	32.877 13.400
	32.925 13.470
	33.004 13.512
	33.115 13.526
	33.225 13.540
	33.302 13.585
	33.344 13.659
	33.353 13.764
	33.360 13.868
	33.397 13.942
	33.464 13.987
	33.560 14.002
	33.657 14.016
	33.726 14.057
	33.768 14.127
	33.782 14.224
	33.796 14.321
	33.838 14.391
	33.907 14.432
	34.004 14.446
	34.101 14.460
	34.171 14.502
	34.213 14.571
	34.227 14.669
	34.236 14.769
	34.266 14.847
	34.385 14.938
	34.462 14.966
	34.532 15.002
	34.655 15.097
	 /
\plot 34.655 15.097 34.766 15.208 /
\linethickness= 0.500pt
\setplotsymbol ({\thinlinefont .})
%
%
%
\plot	34.830 15.272 34.687 15.399
 	34.615 15.456
	34.544 15.502
	34.473 15.536
	34.401 15.558
	34.294 15.633
	34.267 15.703
	34.258 15.796
	34.245 15.886
	34.207 15.950
	34.142 15.989
	34.052 16.002
	33.961 16.015
	33.893 16.054
	33.849 16.118
	33.830 16.208
	33.810 16.300
	33.766 16.367
	33.699 16.411
	33.607 16.431
	33.516 16.452
	33.449 16.502
	33.405 16.579
	33.385 16.685
	33.364 16.789
	33.318 16.863
	33.245 16.908
	33.147 16.923
	33.049 16.936
	32.976 16.974
	32.930 17.039
	32.909 17.129
	32.889 17.220
	32.845 17.288
	32.778 17.332
	32.687 17.351
	32.595 17.371
	32.528 17.415
	32.484 17.482
	32.464 17.574
	32.445 17.665
	32.401 17.732
	32.333 17.776
	32.242 17.796
	32.149 17.813
	32.075 17.847
	31.988 17.971
	31.960 18.047
	31.925 18.117
	31.829 18.240
	 /
\plot 31.829 18.240 31.718 18.351 /
\linethickness= 0.500pt
\setplotsymbol ({\thinlinefont .})
%
%
%
\plot	23.146 15.272 23.289 15.399
 	23.408 15.526
	23.449 15.589
	23.479 15.653
	23.566 15.748
	23.640 15.772
	23.733 15.780
	23.823 15.793
	23.888 15.831
	23.927 15.896
	23.940 15.986
	23.951 16.076
	23.987 16.141
	24.047 16.180
	24.130 16.192
	24.213 16.206
	24.273 16.248
	24.309 16.318
	24.320 16.415
	24.336 16.512
	24.384 16.581
	24.463 16.623
	24.575 16.637
	24.685 16.652
	24.761 16.697
	24.804 16.771
	24.813 16.875
	24.820 16.979
	24.856 17.054
	24.923 17.098
	25.019 17.113
	25.116 17.127
	25.186 17.169
	25.227 17.238
	25.241 17.335
	25.255 17.433
	25.297 17.502
	25.366 17.544
	25.463 17.558
	25.561 17.572
	25.630 17.613
	25.672 17.683
	25.686 17.780
	25.696 17.880
	25.725 17.959
	25.845 18.050
	25.921 18.078
	25.991 18.113
	26.114 18.209
	 /
\plot 26.114 18.209 26.226 18.320 /
\linethickness= 0.500pt
\setplotsymbol ({\thinlinefont .})
%
%
%
\plot	14.637 18.351 14.780 18.224
 	14.851 18.167
	14.922 18.121
	14.994 18.088
	15.065 18.066
	15.173 17.990
	15.199 17.920
	15.208 17.828
	15.221 17.737
	15.260 17.673
	15.324 17.634
	15.415 17.621
	15.506 17.608
	15.573 17.570
	15.617 17.505
	15.637 17.415
	15.657 17.324
	15.700 17.256
	15.768 17.212
	15.859 17.193
	15.950 17.171
	16.018 17.121
	16.062 17.044
	16.081 16.939
	16.102 16.834
	16.149 16.760
	16.221 16.715
	16.320 16.701
	16.418 16.688
	16.490 16.649
	16.537 16.584
	16.558 16.494
	16.577 16.403
	16.621 16.335
	16.689 16.292
	16.780 16.272
	16.871 16.252
	16.939 16.208
	16.982 16.141
	17.002 16.050
	17.022 15.958
	17.066 15.891
	17.133 15.847
	17.224 15.827
	17.318 15.811
	17.391 15.776
	17.478 15.653
	17.506 15.576
	17.542 15.506
	17.637 15.383
	 /
\plot 17.637 15.383 17.748 15.272 /
\linethickness= 0.500pt
\setplotsymbol ({\thinlinefont .})
%
%
%
\plot	 5.048 15.113  5.048 14.954
 	 5.053 14.875
	 5.068 14.796
	 5.093 14.716
	 5.128 14.637
	 5.152 14.557
	 5.147 14.478
	 5.113 14.399
	 5.048 14.319
	 4.989 14.240
	 4.969 14.161
	 4.989 14.081
	 5.048 14.002
	 5.108 13.922
	 5.128 13.843
	 5.108 13.764
	 5.048 13.684
	 4.989 13.605
	 4.969 13.526
	 4.989 13.446
	 5.048 13.367
	 5.108 13.287
	 5.128 13.208
	 5.108 13.129
	 5.048 13.049
	 4.989 12.970
	 4.969 12.891
	 4.989 12.811
	 5.048 12.732
	 5.108 12.652
	 5.128 12.573
	 5.108 12.494
	 5.048 12.414
	 4.989 12.335
	 4.969 12.256
	 4.989 12.176
	 5.048 12.097
	 5.113 12.017
	 5.147 11.938
	 5.152 11.859
	 5.128 11.779
	 5.093 11.700
	 5.068 11.620
	 5.053 11.541
	 5.048 11.462
	 /
\plot  5.048 11.462  5.048 11.303 /
\linethickness= 0.500pt
\setplotsymbol ({\thinlinefont .})
%
%
%
\plot	 5.080 15.113  4.921 15.240
 	 4.841 15.298
	 4.759 15.343
	 4.674 15.377
	 4.588 15.399
	 4.511 15.427
	 4.457 15.478
	 4.424 15.554
	 4.413 15.653
	 4.399 15.753
	 4.358 15.831
	 4.288 15.888
	 4.191 15.923
	 4.093 15.955
	 4.020 16.006
	 3.974 16.074
	 3.953 16.161
	 3.932 16.244
	 3.885 16.304
	 3.813 16.339
	 3.715 16.351
	 3.618 16.366
	 3.548 16.411
	 3.506 16.485
	 3.493 16.589
	 3.479 16.693
	 3.437 16.764
	 3.367 16.804
	 3.270 16.812
	 3.173 16.819
	 3.104 16.855
	 3.062 16.922
	 3.048 17.018
	 3.036 17.115
	 3.000 17.185
	 2.941 17.226
	 2.857 17.240
	 2.774 17.254
	 2.715 17.296
	 2.679 17.365
	 2.667 17.462
	 2.653 17.560
	 2.611 17.629
	 2.542 17.671
	 2.445 17.685
	 2.344 17.696
	 2.262 17.728
	 2.201 17.783
	 2.159 17.859
	 2.124 17.944
	 2.084 18.022
	 2.037 18.095
	 1.984 18.161
	 /
\plot  1.984 18.161  1.873 18.288 /
\linethickness= 0.500pt
\setplotsymbol ({\thinlinefont .})
%
%
%
\plot	 8.223 18.288  8.096 18.129
 	 8.039 18.049
	 7.993 17.967
	 7.959 17.882
	 7.938 17.796
	 7.910 17.719
	 7.858 17.665
	 7.783 17.632
	 7.683 17.621
	 7.583 17.607
	 7.505 17.566
	 7.448 17.496
	 7.414 17.399
	 7.381 17.301
	 7.330 17.228
	 7.262 17.182
	 7.176 17.161
	 7.092 17.140
	 7.033 17.093
	 6.997 17.021
	 6.985 16.923
	 6.970 16.826
	 6.925 16.756
	 6.851 16.714
	 6.747 16.701
	 6.644 16.687
	 6.572 16.645
	 6.533 16.575
	 6.525 16.478
	 6.518 16.381
	 6.481 16.312
	 6.414 16.270
	 6.318 16.256
	 6.221 16.244
	 6.152 16.208
	 6.110 16.149
	 6.096 16.066
	 6.082 15.982
	 6.040 15.923
	 5.971 15.887
	 5.874 15.875
	 5.777 15.861
	 5.707 15.819
	 5.665 15.750
	 5.652 15.653
	 5.641 15.552
	 5.608 15.470
	 5.553 15.409
	 5.477 15.367
	 5.393 15.332
	 5.314 15.292
	 5.242 15.245
	 5.175 15.192
	 /
\plot  5.175 15.192  5.048 15.081 /
%
%
\put{\SetFigFont{9}{10.8}{it}k} [lB] at  4.731 10.192
%
%
\put{\SetFigFont{9}{10.8}{it}q} [lB] at 20.130 16.066
%
%
\put{\SetFigFont{9}{10.8}{it}k} [lB] at 13.938 11.144
%
%
\put{\SetFigFont{9}{10.8}{it}k} [lB] at 31.083 11.303
%
%
\put{\SetFigFont{7}{8.4}{it}1} [lB] at 31.560 11.144
%
%
\put{\SetFigFont{7}{8.4}{it}1} [lB] at 14.415 10.986
%
%
\put{\SetFigFont{7}{8.4}{it}1} [lB] at  5.207  9.874
%
%
\put{\SetFigFont{9}{10.8}{rm}(a)} [lB] at  4.413  8.287
%
%
\put{\SetFigFont{9}{10.8}{rm}(b)} [lB] at 19.653  8.287
%
%
\put{\SetFigFont{9}{10.8}{rm}(c)} [lB] at 34.258  8.287
%
%
\put{\SetFigFont{9}{10.8}{it}k} [lB] at 13.780 18.923
%
%
\put{\SetFigFont{7}{8.4}{it}2} [lB] at 14.256 18.764
%
%
\put{\SetFigFont{9}{10.8}{it}k} [lB] at 31.083 18.923
%
%
\put{\SetFigFont{7}{8.4}{it}2} [lB] at 31.560 18.764
%
%
\put{\SetFigFont{9}{10.8}{it}k} [lB] at 26.162 18.923
%
%
\put{\SetFigFont{7}{8.4}{it}3} [lB] at 26.638 18.764
%
%
\put{\SetFigFont{9}{10.8}{it}k} [lB] at 26.321 11.144
%
%
\put{\SetFigFont{7}{8.4}{it}4} [lB] at 26.797 10.986
%
%
\put{\SetFigFont{7}{8.4}{it}4} [lB] at 38.544 10.986
%
%
\put{\SetFigFont{9}{10.8}{it}k} [lB] at 37.910 18.923
%
%
\put{\SetFigFont{7}{8.4}{it}3} [lB] at 38.386 18.764
%
%
\put{\SetFigFont{9}{10.8}{it}k} [lB] at 38.068 11.144
%
%
\put{\SetFigFont{7}{8.4}{it}2} [lB] at  1.556 18.605
%
%
\put{\SetFigFont{9}{10.8}{it}k} [lB] at  1.079 18.764
%
%
\put{\SetFigFont{9}{10.8}{it}k} [lB] at  8.223 18.764
%
%
\put{\SetFigFont{7}{8.4}{it}3} [lB] at  8.700 18.605
\linethickness=0pt
\putrectangle corners at  1.079 20.098 and 38.544  8.287
\endpicture}

\caption[f1]{\label{f1}{Basic tree diagrams involving gauge particles.
All momenta are inwards with $\sum k_i=0$.}}
\end{figure}
\bigskip

We proceed by imposing the condition that the gluon tree amplitudes
should be invariant under the Abelian gauge transformation given by
(\ref{eq2.4}). This property \cite{feynman} follows in consequence of the fact
that the external lines satisfy the free equation of motion (\ref{eq2.3}).
We use this constraint on the 3-gluon vertex shown in Fig. 1a and perform
a gauge transformation on the field $A^a_\alpha\left(k_1\right)$. Since
the trilinear gluon coupling proportional to $g\, f_{abc}$ satisfies
identically the above constraint, when we make use of momentum conservation,
we find the condition that
\begin{equation}\label{eq2.6}
e_0\left({k_2}_\beta\ {k_3}_\gamma - k_2\cdot k_3 \eta_{\beta\gamma}\right)
\omega^a\,d_{abc} A^b_\beta(k_2) A^c_\gamma(k_3)=0.
\end{equation}
Because $k_2$ and $k_3$ are arbitrary and independent momenta, this equation
requires the vanishing of the coupling constant $e_0$
\begin{equation}\label{eq2.7}
e_0=0.
\end{equation}
Therefore, in this case the Abelian gauge invariance determines basically
the structure of the trilinear vertex. As we shall see, this special feature
does not occur in the gravity case, which is much more complicated
algebraically.

We now evaluate the contributions from the graph in Fig. 1b and its
permutations to the gluon-gluon scattering amplitude.
In order to perform these calculations, it is simpler to use the Feynman
propagator
$\eta_{\mu\nu}/q^2$. In view of the Abelian gauge invariance
of this amplitude, we must equate the negative of the
gauge variation of these contributions
to the corresponding variations associated with the 4-gluon vertex shown
in Fig 1c. Then, under a gauge transformation
of the gluon field $A^a_\alpha(k_1)$, we find that
\begin{eqnarray}\label{eq2.8}
\left[\delta\;tree\right]_{1c}=-g^2 & \left[f_{abe}f_{cde}\left(
{k_1}_\beta \eta_{\sigma\gamma}+
{k_1}_\sigma \eta_{\beta\gamma}-
2 {k_1}_\gamma \eta_{\beta\sigma}\right)+\right.\nonumber\\
& \left. f_{ace}f_{bde}\left(
{k_1}_\gamma \eta_{\sigma\beta}+
{k_1}_\sigma \eta_{\beta\gamma}-
2 {k_1}_\beta \eta_{\gamma\sigma}\right)\right]\nonumber \\
& \times\omega^a A^b_\beta(k_2) A^c_\gamma(k_3) A^d_\sigma(k_4),
\end{eqnarray}
where we have used the Jacobi identity
\begin{equation}\label{eq2.9}
f_{abe}f_{cde}+f_{ace}f_{dbe}+f_{ade}f_{bce}=0
\end{equation}
to eliminate contributions proportional to $f_{ade}f_{bce}$.

\begin{figure}
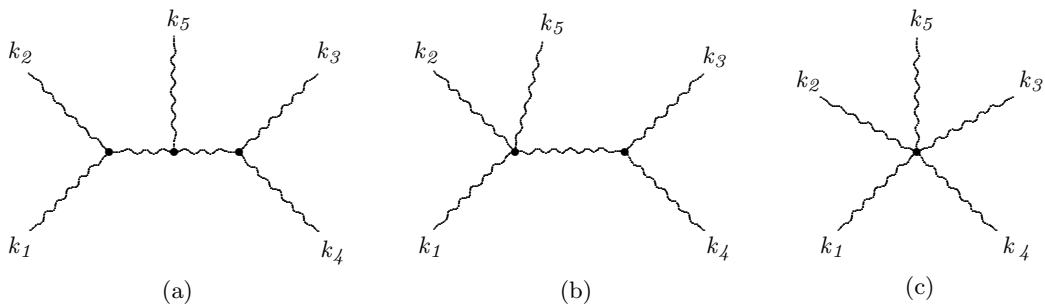

\font\thinlinefont=cmr5
\begingroup\makeatletter\ifx\SetFigFont\undefined
\def\x#1#2#3#4#5#6#7\relax{\def\x{#1#2#3#4#5#6}}%
\expandafter\x\fmtname xxxxxx\relax \def\y{splain}%
\ifx\x\y   
\gdef\SetFigFont#1#2#3{%
  \ifnum #1<17\tiny\else \ifnum #1<20\small\else
  \ifnum #1<24\normalsize\else \ifnum #1<29\large\else
  \ifnum #1<34\Large\else \ifnum #1<41\LARGE\else
     \huge\fi\fi\fi\fi\fi\fi
  \csname #3\endcsname}%
\else
\gdef\SetFigFont#1#2#3{\begingroup
  \count@#1\relax \ifnum 25<\count@\count@25\fi
  \def\x{\endgroup\@setsize\SetFigFont{#2pt}}%
  \expandafter\x
    \csname \romannumeral\the\count@ pt\expandafter\endcsname
    \csname @\romannumeral\the\count@ pt\endcsname
  \csname #3\endcsname}%
\fi
\fi\endgroup
\mbox{\beginpicture
\setcoordinatesystem units < 0.340cm, 0.340cm>
\unitlength= 0.340cm
\linethickness=1pt
\setplotsymbol ({\makebox(0,0)[l]{\tencirc\symbol{'160}}})
\setshadesymbol ({\thinlinefont .})
\setlinear
%
%
\linethickness= 0.500pt
\setplotsymbol ({\thinlinefont .})
\put{\makebox(0,0)[l]{\circle*{ 0.318}}}
at 27.908 15.272
%
\linethickness= 0.500pt
\setplotsymbol ({\thinlinefont .})
\put{\makebox(0,0)[l]{\circle*{ 0.318}}}
at 10.287 15.272
%
\linethickness= 0.500pt
\setplotsymbol ({\thinlinefont .})
\put{\makebox(0,0)[l]{\circle*{ 0.318}}}
at 39.338 15.272
%
\linethickness= 0.500pt
\setplotsymbol ({\thinlinefont .})
\put{\makebox(0,0)[l]{\circle*{ 0.318}}}
at 12.827 15.272
%
\linethickness= 0.500pt
\setplotsymbol ({\thinlinefont .})
\put{\makebox(0,0)[l]{\circle*{ 0.318}}}
at  7.747 15.272
%
\linethickness= 0.500pt
\setplotsymbol ({\thinlinefont .})
\put{\makebox(0,0)[l]{\circle*{ 0.318}}}
at 23.622 15.272
\linethickness= 0.500pt
\setplotsymbol ({\thinlinefont .})
%
%
%
\plot	31.020 12.192 30.877 12.319
 	30.805 12.377
	30.734 12.422
	30.663 12.456
	30.591 12.478
	30.484 12.553
	30.457 12.624
	30.448 12.716
	30.435 12.806
	30.397 12.871
	30.332 12.909
	30.242 12.922
	30.151 12.935
	30.083 12.974
	30.039 13.038
	30.020 13.129
	30.000 13.220
	29.956 13.287
	29.889 13.331
	29.797 13.351
	29.706 13.373
	29.639 13.422
	29.595 13.500
	29.575 13.605
	29.554 13.709
	29.508 13.783
	29.435 13.828
	29.337 13.843
	29.239 13.856
	29.166 13.895
	29.120 13.959
	29.099 14.049
	29.079 14.141
	29.035 14.208
	28.968 14.252
	28.877 14.272
	28.785 14.291
	28.718 14.335
	28.674 14.403
	28.654 14.494
	28.635 14.585
	28.591 14.653
	28.523 14.696
	28.432 14.716
	28.339 14.733
	28.265 14.768
	28.178 14.891
	28.150 14.967
	28.115 15.038
	28.019 15.161
	 /
\plot 28.019 15.161 27.908 15.272 /
\linethickness= 0.500pt
\setplotsymbol ({\thinlinefont .})
%
%
%
\plot	27.908 15.272 27.718 15.288
 	27.627 15.292
	27.547 15.288
	27.477 15.276
	27.416 15.256
	27.293 15.256
	27.223 15.292
	27.146 15.351
	27.068 15.405
	26.991 15.423
	26.917 15.405
	26.845 15.351
	26.774 15.297
	26.706 15.276
	26.639 15.289
	26.575 15.335
	26.509 15.381
	26.440 15.391
	26.366 15.365
	26.289 15.304
	26.209 15.245
	26.126 15.228
	26.042 15.253
	25.956 15.319
	25.869 15.385
	25.785 15.407
	25.703 15.385
	25.622 15.319
	25.543 15.253
	25.463 15.228
	25.384 15.245
	25.305 15.304
	25.225 15.363
	25.146 15.383
	25.067 15.363
	24.987 15.304
	24.908 15.244
	24.829 15.224
	24.749 15.244
	24.670 15.304
	24.590 15.363
	24.511 15.383
	24.432 15.363
	24.352 15.304
	24.274 15.239
	24.197 15.204
	24.123 15.199
	24.051 15.224
	23.978 15.259
	23.904 15.284
	23.827 15.299
	23.749 15.304
	 /
\plot 23.749 15.304 23.590 15.304 /
\linethickness= 0.500pt
\setplotsymbol ({\thinlinefont .})
%
%
%
\plot	23.622 15.272 23.654 15.478
 	23.666 15.576
	23.670 15.665
	23.666 15.743
	23.654 15.811
	23.674 15.935
	23.797 16.050
	23.867 16.110
	23.904 16.181
	23.907 16.261
	23.876 16.351
	23.844 16.442
	23.844 16.522
	23.876 16.592
	23.940 16.653
	24.039 16.780
	24.044 16.855
	24.019 16.939
	23.996 17.025
	24.007 17.109
	24.052 17.192
	24.130 17.272
	24.206 17.353
	24.245 17.439
	24.246 17.528
	24.209 17.621
	24.172 17.711
	24.170 17.788
	24.273 17.907
	24.380 18.030
	24.383 18.112
	24.352 18.209
	24.321 18.305
	24.324 18.387
	24.432 18.510
	24.539 18.633
	24.542 18.716
	24.511 18.812
	24.476 18.909
	24.467 18.994
	24.484 19.068
	24.527 19.129
	24.618 19.252
	24.649 19.323
	24.670 19.399
	 /
\plot 24.670 19.399 24.701 19.558 /
\linethickness= 0.500pt
\setplotsymbol ({\thinlinefont .})
%
%
%
\plot	39.338 19.717 39.322 19.479
 	39.321 19.367
	39.334 19.272
	39.361 19.193
	39.402 19.129
	39.430 19.010
	39.393 18.945
	39.322 18.875
	39.255 18.802
	39.227 18.725
	39.239 18.643
	39.291 18.558
	39.346 18.473
	39.370 18.391
	39.362 18.314
	39.322 18.240
	39.283 18.169
	39.275 18.098
	39.299 18.026
	39.354 17.955
	39.407 17.881
	39.422 17.804
	39.399 17.722
	39.338 17.637
	39.278 17.551
	39.255 17.466
	39.270 17.384
	39.322 17.304
	39.376 17.225
	39.394 17.149
	39.376 17.075
	39.322 17.002
	39.269 16.927
	39.251 16.843
	39.269 16.752
	39.322 16.653
	39.377 16.555
	39.398 16.466
	39.385 16.388
	39.338 16.320
	39.291 16.253
	39.275 16.181
	39.291 16.102
	39.338 16.018
	39.389 15.933
	39.414 15.851
	39.413 15.774
	39.386 15.700
	39.352 15.629
	39.330 15.558
	39.320 15.486
	39.322 15.415
	 /
\plot 39.322 15.415 39.338 15.272 /
\linethickness= 0.500pt
\setplotsymbol ({\thinlinefont .})
%
%
%
\plot	43.117 17.431 42.942 17.351
 	42.861 17.309
	42.791 17.260
	42.688 17.145
	42.577 17.066
	42.497 17.062
	42.402 17.081
	42.309 17.097
	42.235 17.081
	42.148 16.954
	42.116 16.874
	42.069 16.824
	42.005 16.803
	41.926 16.812
	41.847 16.818
	41.783 16.788
	41.735 16.722
	41.704 16.621
	41.668 16.522
	41.608 16.462
	41.525 16.443
	41.418 16.462
	41.310 16.482
	41.223 16.462
	41.159 16.403
	41.116 16.304
	41.075 16.203
	41.013 16.141
	40.932 16.116
	40.831 16.129
	40.730 16.144
	40.652 16.125
	40.595 16.072
	40.561 15.986
	40.527 15.899
	40.473 15.843
	40.400 15.819
	40.307 15.827
	40.212 15.835
	40.136 15.811
	40.077 15.756
	40.037 15.669
	39.999 15.577
	39.949 15.510
	39.888 15.466
	39.815 15.446
	39.735 15.434
	39.656 15.415
	39.576 15.387
	39.497 15.351
	 /
\plot 39.497 15.351 39.338 15.272 /
\linethickness= 0.500pt
\setplotsymbol ({\thinlinefont .})
%
%
%
\plot	39.338 15.272 39.180 15.351
 	39.100 15.386
	39.021 15.411
	38.941 15.426
	38.862 15.431
	38.790 15.442
	38.731 15.478
	38.686 15.538
	38.656 15.621
	38.622 15.701
	38.568 15.752
	38.495 15.773
	38.402 15.764
	38.308 15.755
	38.235 15.776
	38.181 15.826
	38.148 15.907
	38.113 15.988
	38.056 16.042
	37.978 16.067
	37.878 16.066
	37.777 16.066
	37.695 16.097
	37.634 16.161
	37.592 16.256
	37.551 16.348
	37.493 16.403
	37.416 16.420
	37.322 16.399
	37.227 16.378
	37.148 16.395
	37.084 16.449
	37.036 16.542
	36.989 16.638
	36.925 16.704
	36.846 16.741
	36.751 16.748
	36.656 16.752
	36.580 16.780
	36.481 16.907
	36.382 17.030
	36.305 17.053
	36.211 17.050
	36.116 17.044
	36.036 17.058
	35.925 17.145
	35.830 17.260
	35.770 17.309
	35.703 17.351
	 /
\plot 35.703 17.351 35.560 17.431 /
\linethickness= 0.500pt
\setplotsymbol ({\thinlinefont .})
%
%
%
\plot	42.450 12.192 42.307 12.319
 	42.235 12.377
	42.164 12.422
	42.093 12.456
	42.021 12.478
	41.914 12.553
	41.887 12.624
	41.878 12.716
	41.865 12.806
	41.827 12.871
	41.762 12.909
	41.672 12.922
	41.581 12.935
	41.513 12.974
	41.469 13.038
	41.450 13.129
	41.430 13.220
	41.386 13.287
	41.319 13.331
	41.227 13.351
	41.136 13.373
	41.069 13.422
	41.025 13.500
	41.005 13.605
	40.984 13.709
	40.938 13.783
	40.865 13.828
	40.767 13.843
	40.669 13.856
	40.596 13.895
	40.550 13.959
	40.529 14.049
	40.509 14.141
	40.465 14.208
	40.398 14.252
	40.307 14.272
	40.215 14.291
	40.148 14.335
	40.104 14.403
	40.084 14.494
	40.065 14.585
	40.021 14.653
	39.953 14.696
	39.862 14.716
	39.769 14.733
	39.695 14.768
	39.608 14.891
	39.580 14.967
	39.545 15.038
	39.449 15.161
	 /
\plot 39.449 15.161 39.338 15.272 /
\linethickness= 0.500pt
\setplotsymbol ({\thinlinefont .})
%
%
%
\plot	39.338 15.272 39.195 15.145
 	39.130 15.081
	39.076 15.018
	39.005 14.891
	38.918 14.796
	38.844 14.772
	38.751 14.764
	38.661 14.751
	38.596 14.712
	38.557 14.648
	38.544 14.557
	38.533 14.467
	38.497 14.403
	38.437 14.364
	38.354 14.351
	38.271 14.337
	38.211 14.295
	38.175 14.226
	38.163 14.129
	38.148 14.032
	38.100 13.962
	38.021 13.920
	37.910 13.906
	37.799 13.892
	37.723 13.847
	37.680 13.773
	37.671 13.668
	37.664 13.564
	37.628 13.490
	37.561 13.445
	37.465 13.430
	37.368 13.416
	37.298 13.375
	37.257 13.305
	37.243 13.208
	37.229 13.111
	37.187 13.041
	37.118 13.000
	37.020 12.986
	36.923 12.972
	36.854 12.930
	36.812 12.861
	36.798 12.764
	36.788 12.663
	36.759 12.585
	36.640 12.494
	36.563 12.466
	36.493 12.430
	36.370 12.335
	 /
\plot 36.370 12.335 36.258 12.224 /
\linethickness= 0.500pt
\setplotsymbol ({\thinlinefont .})
%
%
\plot	10.128 15.272 10.287 15.272
 	10.366 15.277
	10.446 15.292
	10.525 15.316
	10.604 15.351
	10.684 15.376
	10.763 15.371
	10.843 15.336
	10.922 15.272
	11.001 15.212
	11.081 15.192
	11.160 15.212
	11.239 15.272
	11.319 15.331
	11.398 15.351
	11.478 15.331
	11.557 15.272
	11.636 15.212
	11.716 15.192
	11.795 15.212
	11.874 15.272
	11.954 15.336
	12.033 15.371
	12.113 15.376
	12.192 15.351
	12.271 15.316
	12.351 15.292
	12.430 15.277
	12.509 15.272
	 /
\plot 12.509 15.272 12.668 15.272 /
\linethickness= 0.500pt
\setplotsymbol ({\thinlinefont .})
%
%
%
\plot	15.939 12.160 15.796 12.287
 	15.724 12.345
	15.653 12.390
	15.581 12.424
	15.510 12.446
	15.403 12.521
	15.376 12.592
	15.367 12.684
	15.354 12.774
	15.315 12.839
	15.251 12.878
	15.161 12.891
	15.069 12.903
	15.002 12.942
	14.958 13.007
	14.938 13.097
	14.919 13.188
	14.875 13.256
	14.807 13.299
	14.716 13.319
	14.625 13.341
	14.557 13.391
	14.514 13.468
	14.494 13.573
	14.473 13.677
	14.426 13.752
	14.354 13.796
	14.256 13.811
	14.158 13.824
	14.085 13.863
	14.038 13.927
	14.018 14.018
	13.998 14.109
	13.954 14.176
	13.887 14.220
	13.795 14.240
	13.704 14.260
	13.637 14.303
	13.593 14.371
	13.573 14.462
	13.553 14.553
	13.510 14.621
	13.442 14.665
	13.351 14.684
	13.258 14.701
	13.184 14.736
	13.097 14.859
	13.069 14.935
	13.033 15.006
	12.938 15.129
	 /
\plot 12.938 15.129 12.827 15.240 /
\linethickness= 0.500pt
\setplotsymbol ({\thinlinefont .})
%
%
%
\plot	 7.747 15.240  7.604 15.113
 	 7.485 14.986
	 7.443 14.923
	 7.414 14.859
	 7.326 14.764
	 7.253 14.740
	 7.160 14.732
	 7.069 14.719
	 7.005 14.680
	 6.966 14.616
	 6.953 14.526
	 6.941 14.435
	 6.906 14.371
	 6.846 14.332
	 6.763 14.319
	 6.679 14.305
	 6.620 14.264
	 6.584 14.194
	 6.572 14.097
	 6.556 14.000
	 6.509 13.930
	 6.429 13.889
	 6.318 13.875
	 6.208 13.860
	 6.132 13.815
	 6.089 13.741
	 6.080 13.637
	 6.073 13.532
	 6.036 13.458
	 5.970 13.413
	 5.874 13.399
	 5.777 13.385
	 5.707 13.343
	 5.665 13.273
	 5.652 13.176
	 5.638 13.079
	 5.596 13.010
	 5.526 12.968
	 5.429 12.954
	 5.332 12.940
	 5.263 12.898
	 5.221 12.829
	 5.207 12.732
	 5.197 12.632
	 5.167 12.553
	 5.048 12.462
	 4.972 12.434
	 4.901 12.398
	 4.778 12.303
	 /
\plot  4.778 12.303  4.667 12.192 /
\linethickness= 0.500pt
\setplotsymbol ({\thinlinefont .})
%
%
%
\plot	10.287 19.780 10.287 19.590
 	10.290 19.498
	10.299 19.415
	10.314 19.340
	10.335 19.272
	10.347 19.208
	10.335 19.141
	10.299 19.073
	10.239 19.002
	10.186 18.931
	10.168 18.860
	10.186 18.788
	10.239 18.717
	10.294 18.645
	10.315 18.574
	10.302 18.502
	10.255 18.431
	10.209 18.358
	10.196 18.284
	10.217 18.208
	10.271 18.129
	10.323 18.049
	10.335 17.967
	10.307 17.882
	10.239 17.796
	10.174 17.710
	10.152 17.625
	10.174 17.543
	10.239 17.462
	10.307 17.383
	10.335 17.304
	10.323 17.224
	10.271 17.145
	10.218 17.066
	10.200 16.986
	10.218 16.907
	10.271 16.828
	10.325 16.748
	10.343 16.669
	10.325 16.589
	10.271 16.510
	10.218 16.432
	10.200 16.355
	10.218 16.281
	10.271 16.208
	10.330 16.136
	10.362 16.062
	10.369 15.985
	10.350 15.907
	10.323 15.827
	10.303 15.748
	10.291 15.669
	10.287 15.589
	 /
\plot 10.287 15.589 10.287 15.431 /
\linethickness= 0.500pt
\setplotsymbol ({\thinlinefont .})
%
%
%
\plot	 4.604 18.351  4.747 18.224
 	 4.816 18.167
	 4.882 18.121
	 5.001 18.066
	 5.092 17.986
	 5.123 17.911
	 5.144 17.812
	 5.169 17.714
	 5.215 17.641
	 5.280 17.594
	 5.366 17.574
	 5.450 17.556
	 5.513 17.518
	 5.553 17.461
	 5.572 17.383
	 5.591 17.304
	 5.632 17.240
	 5.694 17.193
	 5.779 17.161
	 5.865 17.126
	 5.933 17.070
	 5.984 16.991
	 6.017 16.891
	 6.050 16.793
	 6.104 16.720
	 6.177 16.674
	 6.271 16.653
	 6.362 16.635
	 6.429 16.597
	 6.473 16.540
	 6.493 16.462
	 6.512 16.383
	 6.552 16.319
	 6.615 16.272
	 6.699 16.240
	 6.786 16.207
	 6.854 16.157
	 6.905 16.088
	 6.937 16.002
	 6.969 15.919
	 7.017 15.859
	 7.080 15.823
	 7.160 15.811
	 7.241 15.802
	 7.310 15.772
	 7.414 15.653
	 7.456 15.576
	 7.505 15.506
	 7.620 15.383
	 /
\plot  7.620 15.383  7.747 15.272 /
\linethickness= 0.500pt
\setplotsymbol ({\thinlinefont .})
%
%
%
\plot	12.827 15.272 12.970 15.399
 	13.089 15.526
	13.131 15.589
	13.160 15.653
	13.248 15.748
	13.321 15.772
	13.414 15.780
	13.505 15.793
	13.569 15.831
	13.608 15.896
	13.621 15.986
	13.633 16.076
	13.668 16.141
	13.728 16.180
	13.811 16.192
	13.895 16.206
	13.954 16.248
	13.990 16.318
	14.002 16.415
	14.018 16.512
	14.065 16.581
	14.145 16.623
	14.256 16.637
	14.366 16.652
	14.442 16.697
	14.485 16.771
	14.494 16.875
	14.501 16.979
	14.538 17.054
	14.604 17.098
	14.700 17.113
	14.797 17.127
	14.867 17.169
	14.909 17.238
	14.922 17.335
	14.936 17.433
	14.978 17.502
	15.048 17.544
	15.145 17.558
	15.242 17.572
	15.311 17.613
	15.353 17.683
	15.367 17.780
	15.377 17.880
	15.407 17.959
	15.526 18.050
	15.602 18.078
	15.673 18.113
	15.796 18.209
	 /
\plot 15.796 18.209 15.907 18.320 /
\linethickness= 0.500pt
\setplotsymbol ({\thinlinefont .})
%
%
\plot	 7.906 15.272  8.064 15.272
 	 8.144 15.277
	 8.223 15.292
	 8.303 15.316
	 8.382 15.351
	 8.461 15.376
	 8.541 15.371
	 8.620 15.336
	 8.700 15.272
	 8.779 15.212
	 8.858 15.192
	 8.938 15.212
	 9.017 15.272
	 9.096 15.331
	 9.176 15.351
	 9.255 15.331
	 9.335 15.272
	 9.414 15.212
	 9.493 15.192
	 9.573 15.212
	 9.652 15.272
	 9.731 15.336
	 9.811 15.371
	 9.890 15.376
	 9.970 15.351
	10.049 15.316
	10.128 15.292
	10.208 15.277
	10.287 15.272
	 /
\plot 10.287 15.272 10.446 15.272 /
\linethickness= 0.500pt
\setplotsymbol ({\thinlinefont .})
%
%
%
\plot	23.527 15.304 23.384 15.176
 	23.318 15.113
	23.265 15.049
	23.193 14.922
	23.106 14.827
	23.033 14.803
	22.939 14.796
	22.849 14.783
	22.785 14.744
	22.746 14.679
	22.733 14.589
	22.721 14.499
	22.685 14.434
	22.626 14.396
	22.543 14.383
	22.459 14.369
	22.400 14.327
	22.364 14.258
	22.352 14.160
	22.336 14.063
	22.289 13.994
	22.209 13.952
	22.098 13.938
	21.988 13.923
	21.911 13.879
	21.869 13.804
	21.860 13.700
	21.853 13.596
	21.816 13.522
	21.750 13.477
	21.654 13.462
	21.556 13.448
	21.487 13.406
	21.445 13.337
	21.431 13.240
	21.417 13.143
	21.376 13.073
	21.306 13.031
	21.209 13.018
	21.112 13.004
	21.042 12.962
	21.001 12.892
	20.987 12.795
	20.977 12.695
	20.947 12.617
	20.828 12.525
	20.752 12.498
	20.681 12.462
	20.558 12.367
	 /
\plot 20.558 12.367 20.447 12.256 /
\linethickness= 0.500pt
\setplotsymbol ({\thinlinefont .})
%
%
%
\plot	20.447 18.415 20.590 18.288
 	20.661 18.230
	20.733 18.185
	20.804 18.151
	20.876 18.129
	20.983 18.054
	21.010 17.983
	21.018 17.891
	21.031 17.801
	21.070 17.736
	21.135 17.698
	21.225 17.685
	21.316 17.672
	21.384 17.633
	21.427 17.569
	21.447 17.478
	21.467 17.387
	21.511 17.320
	21.578 17.276
	21.669 17.256
	21.761 17.234
	21.828 17.185
	21.872 17.107
	21.892 17.002
	21.912 16.898
	21.959 16.824
	22.032 16.779
	22.130 16.764
	22.228 16.751
	22.300 16.712
	22.347 16.648
	22.368 16.558
	22.388 16.466
	22.431 16.399
	22.499 16.355
	22.590 16.335
	22.681 16.316
	22.749 16.272
	22.793 16.204
	22.812 16.113
	22.832 16.022
	22.876 15.954
	22.943 15.911
	23.035 15.891
	23.128 15.874
	23.201 15.839
	23.289 15.716
	23.316 15.640
	23.352 15.569
	23.447 15.446
	 /
\plot 23.447 15.446 23.559 15.335 /
\linethickness= 0.500pt
\setplotsymbol ({\thinlinefont .})
%
%
%
\plot	27.908 15.272 28.051 15.399
 	28.170 15.526
	28.212 15.589
	28.242 15.653
	28.329 15.748
	28.402 15.772
	28.496 15.780
	28.586 15.793
	28.650 15.831
	28.689 15.896
	28.702 15.986
	28.714 16.076
	28.750 16.141
	28.809 16.180
	28.893 16.192
	28.976 16.206
	29.035 16.248
	29.071 16.318
	29.083 16.415
	29.099 16.512
	29.146 16.581
	29.226 16.623
	29.337 16.637
	29.447 16.652
	29.524 16.697
	29.566 16.771
	29.575 16.875
	29.582 16.979
	29.619 17.054
	29.685 17.098
	29.782 17.113
	29.879 17.127
	29.948 17.169
	29.990 17.238
	30.004 17.335
	30.018 17.433
	30.059 17.502
	30.129 17.544
	30.226 17.558
	30.323 17.572
	30.393 17.613
	30.434 17.683
	30.448 17.780
	30.458 17.880
	30.488 17.959
	30.607 18.050
	30.683 18.078
	30.754 18.113
	30.877 18.209
	 /
\plot 30.877 18.209 30.988 18.320 /
%
%
\put{\SetFigFont{7}{8.4}{it}1} [lB] at 20.288 11.144
%
%
\put{\SetFigFont{9}{10.8}{it}k} [lB] at 19.812 11.303
%
%
\put{\SetFigFont{9}{10.8}{rm}(a)} [lB] at  9.811  9.557
%
%
\put{\SetFigFont{9}{10.8}{rm}(b)} [lB] at 25.368  9.557
%
%
\put{\SetFigFont{9}{10.8}{rm}(c)} [lB] at 38.862  9.716
%
%
\put{\SetFigFont{9}{10.8}{it}k} [lB] at 42.672 11.303
%
%
\put{\SetFigFont{7}{8.4}{it}4} [lB] at 43.307 11.144
%
%
\put{\SetFigFont{7}{8.4}{it}5} [lB] at 25.051 19.876
%
%
\put{\SetFigFont{9}{10.8}{it}k} [lB] at 24.575 20.034
%
%
\put{\SetFigFont{9}{10.8}{it}k} [lB] at 19.653 18.923
%
%
\put{\SetFigFont{7}{8.4}{it}2} [lB] at 20.130 18.764
%
%
\put{\SetFigFont{9}{10.8}{it}k} [lB] at 30.925 18.764
%
%
\put{\SetFigFont{7}{8.4}{it}3} [lB] at 31.401 18.605
%
%
\put{\SetFigFont{9}{10.8}{it}k} [lB] at 31.083 11.303
%
%
\put{\SetFigFont{7}{8.4}{it}4} [lB] at 31.560 11.144
%
%
\put{\SetFigFont{9}{10.8}{it}k} [lB] at  9.970 20.193
%
%
\put{\SetFigFont{7}{8.4}{it}5} [lB] at 10.446 20.034
%
%
\put{\SetFigFont{7}{8.4}{it}3} [lB] at 16.320 18.764
%
%
\put{\SetFigFont{9}{10.8}{it}k} [lB] at 15.843 18.923
%
%
\put{\SetFigFont{7}{8.4}{it}4} [lB] at 16.478 10.986
%
%
\put{\SetFigFont{9}{10.8}{it}k} [lB] at 16.002 11.144
%
%
\put{\SetFigFont{9}{10.8}{it}k} [lB] at  3.778 18.923
%
%
\put{\SetFigFont{7}{8.4}{it}2} [lB] at  4.255 18.764
%
%
\put{\SetFigFont{9}{10.8}{it}k} [lB] at  3.778 11.303
%
%
\put{\SetFigFont{7}{8.4}{it}1} [lB] at  4.255 11.144
%
%
\put{\SetFigFont{9}{10.8}{it}k} [lB] at 34.576 17.812
%
%
\put{\SetFigFont{7}{8.4}{it}2} [lB] at 35.052 17.653
%
%
\put{\SetFigFont{9}{10.8}{it}k} [lB] at 43.307 17.812
%
%
\put{\SetFigFont{7}{8.4}{it}3} [lB] at 43.783 17.653
%
%
\put{\SetFigFont{7}{8.4}{it}5} [lB] at 39.497 20.034
%
%
\put{\SetFigFont{9}{10.8}{it}k} [lB] at 39.021 20.193
%
%
\put{\SetFigFont{9}{10.8}{it}k} [lB] at 35.528 11.303
%
%
\put{\SetFigFont{7}{8.4}{it}1} [lB] at 36.005 11.144
\linethickness=0pt
\putrectangle corners at  3.778 21.368 and 43.783  9.557
\endpicture}

\caption[f2]{\label{f2}{
Higher order tree amplitude containing nonlinear couplings of gauge particles
}}
\end{figure}
\bigskip

We can now express the gauge variation on the left hand side of
(\ref{eq2.8}) in terms of the parameters introduced in (\ref{eq2.5}).
Using relations like
\begin{equation}\label{eq2.10}
f_{abe}f_{cde}=\frac{2} {N}\left(\delta_{ac}\delta_{bd}-\delta_{ad}\delta_{bc}
\right)+d_{ace}d_{dbe}-d_{ade}d_{bce},
\end{equation}
and identifying the coefficients of the independent structures appearing
in (\ref{eq2.8}), we obtain the following relations:
\begin{eqnarray}\label{eq2.11}
&l_1&=-l_3=l_0-\frac{g^2}{4},\nonumber \\
& &\\
&l_2&=-l_4=\frac{2}{N}\left(l_0-\frac{g^2}{4}\right). \nonumber
\end{eqnarray}
We thus see that the parameters $l_i$ have not been fully determined by the
gauge invariance property of the gluon-gluon scattering amplitude.
However, we can now apply this condition also to the 5-gluon tree amplitude
represented by diagrams like the one shown in figures 2a and 2b. Due to
the absence of direct 5-gluon couplings, and using the equations
(\ref{eq2.11}),
it is straightforward to show that this constraint yields a further relation:
\begin{equation}\label{eq2.12}
l_0=\frac{g^2}{4}.
\end{equation}
Together with (\ref{eq2.11}), this relation implies the vanishing of the
coupling constants $l_i\;(i=1,2,3,4)$. Substituting these results in
equation (\ref{eq2.5}), and using (\ref{eq2.1}) and (\ref{eq2.7}), we arrive
at the well known expression for the Yang-Mills Lagrangian
\begin{eqnarray}\label{eq2.13}
{\cal L}_{YM}(A)=&\frac{1}{4}&
\left(\partial_\beta A^a_\alpha-\partial_\alpha A^a_\beta +
g\,f_{abc} A^b_\alpha A^c_\beta \right)
\nonumber \\ & &
\left(\partial_\beta A^a_\alpha-\partial_\alpha A^a_\beta +
g\,f_{a{b'}{c'}} A^{b'}_\alpha A^{c'}_\beta \right).
\end{eqnarray}

\section{The gravitational field}\label{sec3}

In this case, it is convenient to introduce a symmetric tensor field
$h_{\mu\nu}$ representing the deviation of the metric tensor $g_{\mu\nu}$
from the flat space Minkowski metric $\eta_{\mu\nu}$:
\begin{equation}\label{eq3.0}
g_{\mu\nu}=\eta_{\mu\nu}+\kappa\; h_{\mu\nu},
\end{equation}
where $\kappa$ is the usual gravitational constant. Gauge symmetry and
Lorentz invariance enable us to get the linearized gravitational Lagrangian
\begin{eqnarray}\label{eq3.1}
{\cal L}^2(h)=&\frac{1}{2}& h_{\mu\nu,\alpha}h_{\mu\nu,\alpha}-
               \frac{1}{2}  h_{\mu\mu,\alpha}h_{\nu\nu,\alpha}+
\nonumber \\ & &
                            h_{\mu\mu,\alpha}h_{\alpha\nu,\nu}-
                            h_{\mu\nu,\nu}h_{\mu\alpha,\alpha}\;\;,
\end{eqnarray}
where the index after a comma indicates differentiation. Although
we are not making explicit the distinction between up and down indices,
the Minkowski metric tensor $\eta_{\mu\nu}$ is implicitly present in all
the contractions of pairs of identical indices
(e. g. $h_{\mu\mu}=\eta_{\mu\nu}h_{\mu\nu}$).


It is easy to verify that the above Lagrangian is invariant under the Abelian
gauge transformation
\begin{equation}\label{eq3.2}
h_{\mu\nu}\rightarrow h_{\mu\nu} + \xi_{\mu,\nu}+ \xi_{\nu,\mu}.
\end{equation}
By varying this Lagrangian one obtains in momentum space the equation of motion
satisfied by a free graviton
\begin{equation}\label{eq3.3}
\left(
k^2 \eta_{\alpha\mu} \eta_{\beta\nu} - k_{\mu} k_{\alpha} \eta_{\beta\nu}
                                     - k_{\mu} k_{\beta} \eta_{\alpha\nu}
                                     + k_{\alpha} k_{\beta} \eta_{\mu\nu}
\right) h_{\mu\nu}(k)=0,
\end{equation}
which is invariant under the gauge transformation
\begin{equation}\label{eq3.4}
\delta h_{\mu\nu}(k)= k_\nu \xi_\mu + k_\mu \xi_\nu
\end{equation}
In order to proceed, we need to parametrize the general structure of the
graviton self-interactions, which we assume to involve products of fields
with two derivative indices. The algebraic complexity is now so great that
we have made use of computer algebra to do the calculations. We start
constructing the 3-graviton vertex ${\cal L}^3$ as a sum over all possible
independent trilinear products of fields with two derivative terms. When
we write all possible such products and use Lorentz invariance, we find
an expression involving 16 independent constants $a_i$
\begin{equation}\label{eq3.5}
\begin{array}{llclclc}
{\cal L}^3\left(h\right)=\kappa\left(\right.&
 a_{1}\, h_{{\mu\nu}}\, h_{{\alpha\beta ,\mu}}\, h_{{\nu\alpha ,\beta}} &+&
 a_{2}\, h_{{\mu\nu}}\, h_{{\alpha\alpha ,\mu}}\, h_{{\nu\beta ,\beta}} &+&
 a_{3}\, h_{{\mu\nu}}\, h_{{\mu\alpha ,\alpha}}\, h_{{\nu\beta ,\beta}} &+
\\
&a_{4}\, h_{{\mu\nu}}\, h_{{\mu\alpha ,\nu}}\, h_{{\alpha\beta ,\beta}} &+&
 a_{5}\, h_{{\mu\nu}}\, h_{{\alpha\beta ,\mu}}\, h_{{\alpha\beta ,\nu}} &+&
 a_{6}\, h_{{\mu\nu}}\, h_{{\mu\alpha ,\beta}}\, h_{{\nu\alpha ,\beta}} &+
\\
&a_{7}\, h_{{\mu\mu}}\, h_{{\nu\alpha ,\beta}}\, h_{{\nu\alpha ,\beta}} &+&
 a_{8}\, h_{{\mu\mu}}\, h_{{\nu\alpha ,\nu}}\, h_{{\alpha\beta ,\beta}} &+&
 a_{9}\, h_{{\mu\nu}}\, h_{{\mu\nu ,\alpha}}\, h_{{\alpha\beta ,\beta}} &+
\\
&a_{10}\, h_{{\mu\mu}}\, h_{{\nu\alpha ,\nu}}\, h_{{\beta\beta ,\alpha}} &+&
 a_{11}\, h_{{\mu\nu}}\, h_{{\alpha\alpha ,\beta}}\, h_{{\mu\nu ,\beta}} &+&
 a_{12}\, h_{{\mu\nu}}\, h_{{\alpha\alpha ,\mu}}\, h_{{\beta\beta ,\nu}} &+
\\
&a_{13}\, h_{{\mu\nu}}\, h_{{\mu\alpha ,\nu}}\, h_{{\beta\beta ,\alpha}} &+&
 a_{14}\, h_{{\mu\nu}}\, h_{{\nu\alpha ,\beta}}\, h_{{\mu\beta ,\alpha}} &+&
 a_{15}\, h_{{\mu\mu}}\, h_{{\nu\alpha ,\beta}}\, h_{{\nu\beta ,\alpha}} &+
\\
&a_{16}\, h_{{\mu\mu}}\, h_{{\nu\nu ,\alpha}}\, h_{{\beta\beta ,\alpha}}&
 \left.\right)&{}&{}&{}&.
\end{array}
\end{equation}

The next steps are done in correspondence with the ones in the Yang-Mills
theory. We attempt to determine these constants, using the requirement of
gauge invariance under the transformation (\ref{eq3.4}) of the 3-graviton
vertex associated with Fig. 1a. Using the equation of motion (\ref{eq3.3})
for the free gravitons and momentum conservation, this results in a set of
7 independent equations for the 16 parameters, which yield the following
relations:
\begin{equation}\label{eq3.6}
\begin{array}{lll}
a_{1}&=&a_{14}-6\,a_{8}-6\,a_{15}-8\,a_{16}-8\,a_{10}-4\,a_{11}-4\,a_{9}
\\
a_{2}&=&-6\,a_{8}-8\,a_{15}-8\,a_{16}-8\,a_{10}+a_{13}-
         2\,a_{11}-2\,a_{9}-4\,a_{12}
\\
a_{3}&=&14\,a_{15}+12\,a_{8}+16\,a_{16}+16\,a_{10}-
         2\,a_{13}+4\,a_{11}+4\,a_{9}+4\,a_{12}  - a_{14}
\\
a_{4}&=&-a_{14}+2\,a_{15}-2\,a_{13}
\\
a_{5}&=&3\,a_{8}+3\,a_{15}+4\,a_{16}+4\,a_{10}+2\,a_{11}+2\,a_{9}
\\
a_{6}&=&-2\,a_{15}+2\,a_{13}
\\
2\,a_{7}&=&-{\displaystyle{{3\,a_{8}}}}-
          {\displaystyle{{3\,a_{15}}}}-4\,a_{16}-4\,a_{10}.
\\
\end{array}
\end{equation}
We remark that after inserting (\ref{eq3.6}) into  (\ref{eq3.5}) the resulting
expression is such that the coefficient of $a_{14}$ is a total derivative.

In contrast to the situation in the Yang-Mills theory [see eq. (\ref{eq2.7})]
we see that in this case we do not have enough conditions to determine all the
parameters of the trilinear graviton couplings. All we can do is to express
${\cal L}^3$ as a function of the parameters which appear on
the right hand side of eq. (\ref{eq3.6}), which we denote collectively be
the set $\tilde a\equiv{a_8, \cdots , a_{16}}$.

It is appropriate to
comment here on the possibility of making a local transformation of the
fields so that
\begin{eqnarray}\label{eq3.7a}
h_{\mu\nu}^{\prime}=h_{\mu\nu}+\kappa\left(\right. &A_1& \eta_{\mu\nu}
\left(h_{\alpha\alpha}\right)^2+A_2\eta_{\mu\nu}h_{\alpha\beta}h_{\beta\alpha}+
\nonumber \\
&A_3& h_{\mu\alpha} h_{\nu\alpha}+
A_4 h_{\mu\nu} h_{\alpha\alpha} \left. \right)
+\cdots,
\end{eqnarray}
where $\cdots$ denote terms of higher order in $\kappa$.
Note that the Abelian gauge transformation (\ref{eq3.4}) is the same for
both fields $h_{\mu\nu}$ and $h_{\mu\nu}^\prime$. Since the terms of order
$\kappa$ in (\ref{eq3.7a}) involve 4 arbitrary parameters, it is possible to
make a redefinition of the fields such that the number of independent
parameters in (\ref{eq3.5}) may be reduced from 9 to 5. Even allowing
for this possibility, we see that in contrast to the Yang-Mills case,
there remains a basic indetermination of the trilinear graviton couplings.

Following the analysis done in the Yang-Mills case, we may evaluate the
contributions from the graph 1b and its permutations to the graviton-graviton
tree amplitude, in terms of the parameters present in the set  $\tilde a$.
Since the gauge invariance condition of the physical tree amplitude
should be valid for any gauge-fixing term added onto (\ref{eq3.1}), it will
be convenient to choose this so that the graviton propagator becomes
\cite{feynman}
\begin{equation}\label{eq3.7}
P_{\mu\nu\alpha\beta}(q)=\frac{
\eta_{\mu\alpha}\eta_{\nu\beta}+
\eta_{\mu\beta}\eta_{\nu\alpha}-
\eta_{\mu\nu}\eta_{\alpha\beta}}{2q^2}.
\end{equation}
(We have verified, in the case of the gravitational Compton scattering by
scalar particles, that no additional information is obtained by
considering a more general class of gauges.)
The result of this evaluation, involving quadratic functions of the parameters
$\tilde a_i$ which are excessively long to write down here, will be employed
subsequently.

Next we must parametrize the structure of the 4-graviton vertex ${\cal L}^4$
indicated in Fig 1c, in terms of all possible quadrilinear products of fields
with two derivatives indices. Proceeding in this way, we find
for ${\cal L}^4$ the following expression  involving 43 independent constants:
\begin{equation}\label{eq3.8}
\begin{array}{rcrcrc}
{\cal L}^4\left(h\right)=&\kappa^2\left(\right.&
 \,b_{1}\, h_{{\mu\nu}}\, h_{{\mu\nu}}\,
h_{{\alpha\beta,\rho}}\, h_{{\alpha\rho,\beta}}&+&
 \,b_{2}\, h_{{\mu\mu}}\, h_{{\nu\nu}}\, h_{{\alpha\beta,\rho}}\,
h_{{\alpha\rho,\beta}} &+
\\
\,b_{3}\, h_{{\mu\nu}}\, h_{{\mu\nu}}\, h_{{\alpha\alpha,\beta}}\,
h_{{\rho\rho,\beta}} &+&
 \,b_{4}\, h_{{\mu\nu}}\, h_{{\alpha\mu}}\, h_{{\beta\beta,\alpha}}\,
h_{{\nu\rho,\rho}} &+&
 \,b_{5}\, h_{{\mu\nu}}\, h_{{\mu\nu}}\, h_{{\alpha\beta,\alpha}}\,
h_{{\beta\rho,\rho}} &+
\\
 \,b_{6}\, h_{{\mu\mu}}\, h_{{\nu\nu}}\, h_{{\alpha\beta,\alpha}}\,
h_{{\beta\rho,\rho}} &+&
 \,b_{7}\, h_{{\mu\mu}}\, h_{{\nu\nu}}\, h_{{\alpha\beta,\rho}}\,
h_{{\alpha\beta,\rho}} &+&
 \,b_{8}\, h_{{\mu\mu}}\, h_{{\nu\nu}}\, h_{{\alpha\alpha,\beta}}\,
h_{{\rho\rho,\beta}} &+
\\
 \,b_{9}\, h_{{\mu\nu}}\, h_{{\alpha\mu}}\, h_{{\beta\beta,\rho}}\,
h_{{\alpha\nu,\rho}} &+&
 \,b_{10}\, h_{{\mu\nu}}\, h_{{\alpha\beta}}\, h_{{\alpha\mu,\nu}}\,
h_{{\rho\rho,\beta}} &+&
 \,b_{11}\, h_{{\mu\nu}}\, h_{{\alpha\alpha}}\, h_{{\beta\mu,\beta}}\,
h_{{\nu\rho,\rho}} &+
\\
 \,b_{12}\, h_{{\mu\nu}}\, h_{{\alpha\mu}}\, h_{{\beta\nu,\beta}}\,
h_{{\alpha\rho,\rho}} &+&
 \,b_{13}\, h_{{\mu\nu}}\, h_{{\alpha\alpha}}\, h_{{\beta\mu,\nu}}\,
h_{{\beta\rho,\rho}} &+&
 \,b_{14}\, h_{{\mu\nu}}\, h_{{\alpha\alpha}}\, h_{{\beta\mu,\nu}}\,
h_{{\rho\rho,\beta}} &+
\\
 \,b_{15}\, h_{{\mu\nu}}\, h_{{\alpha\mu}}\, h_{{\beta\rho,\beta}}\,
h_{{\alpha\nu,\rho}} &+&
 \,b_{16}\, h_{{\mu\mu}}\, h_{{\alpha\nu}}\, h_{{\alpha\nu,\beta}}\,
h_{{\beta\rho,\rho}} &+&
 \,b_{17}\, h_{{\mu\mu}}\, h_{{\alpha\nu}}\, h_{{\alpha\beta,\rho}}\,
h_{{\nu\rho,\beta}} &+
\\
 \,b_{18}\, h_{{\mu\nu}}\, h_{{\alpha\beta}}\, h_{{\alpha\rho,\mu}}\,
h_{{\nu\rho,\beta}} &+&
 \,b_{19}\, h_{{\mu\nu}}\, h_{{\alpha\mu}}\, h_{{\beta\rho,\nu}}\,
h_{{\beta\rho,\alpha}} &+&
 \,b_{20}\, h_{{\mu\nu}}\, h_{{\alpha\mu}}\, h_{{\beta\rho,\alpha}}\,
h_{{\beta\nu,\rho}} &+
\\
 \,b_{21}\, h_{{\mu\nu}}\, h_{{\alpha\alpha}}\, h_{{\beta\rho,\mu}}\,
h_{{\beta\nu,\rho}} &+&
 \,b_{22}\, h_{{\mu\nu}}\, h_{{\alpha\beta}}\, h_{{\mu\rho,\nu}}\,
h_{{\alpha\rho,\beta}} &+&
 \,b_{23}\, h_{{\mu\nu}}\, h_{{\alpha\alpha}}\, h_{{\beta\rho,\mu}}\,
h_{{\beta\rho,\nu}} &+
\\
 \,b_{24}\, h_{{\mu\nu}}\, h_{{\alpha\beta}}\, h_{{\alpha\mu,\rho}}\,
h_{{\beta\nu,\rho}} &+&
 \,b_{25}\, h_{{\mu\nu}}\, h_{{\alpha\mu}}\, h_{{\beta\nu,\rho}}\,
h_{{\alpha\beta,\rho}} &+&
 \,b_{26}\, h_{{\mu\nu}}\, h_{{\alpha\mu}}\, h_{{\beta\nu,\alpha}}\,
h_{{\beta\rho,\rho}} &+
\\
 \,b_{27}\, h_{{\mu\nu}}\, h_{{\mu\nu}}\, h_{{\alpha\beta,\rho}}\,
h_{{\alpha\beta,\rho}} &+&
 \,b_{28}\, h_{{\mu\nu}}\, h_{{\alpha\beta}}\, h_{{\alpha\beta,\rho}}\,
h_{{\mu\nu,\rho}} &+&
 \,b_{29}\, h_{{\mu\nu}}\, h_{{\alpha\beta}}\, h_{{\alpha\beta,\mu}}\,
h_{{\nu\rho,\rho}} &+
\\
 \,b_{30}\, h_{{\mu\nu}}\, h_{{\alpha\beta}}\, h_{{\alpha\mu,\nu}}\,
h_{{\beta\rho,\rho}} &+&
 \,b_{31}\, h_{{\mu\nu}}\, h_{{\mu\nu}}\, h_{{\alpha\beta,\alpha}}\,
h_{{\rho\rho,\beta}} &+&
 \,b_{32}\, h_{{\mu\nu}}\, h_{{\alpha\mu}}\, h_{{\beta\beta,\nu}}\,
h_{{\rho\rho,\alpha}} &+
\\
 \,b_{33}\, h_{{\mu\nu}}\, h_{{\alpha\mu}}\, h_{{\beta\nu,\rho}}\,
h_{{\alpha\rho,\beta}} &+&
 \,b_{34}\, h_{{\mu\mu}}\, h_{{\alpha\nu}}\, h_{{\beta\beta,\nu}}\,
h_{{\rho\rho,\alpha}} &+&
 \,b_{35}\, h_{{\mu\mu}}\, h_{{\nu\nu}}\, h_{{\alpha\alpha,\beta}}\,
h_{{\beta\rho,\rho}} &+
\\
 \,b_{36}\, h_{{\mu\mu}}\, h_{{\alpha\nu}}\, h_{{\beta\beta,\nu}}\,
h_{{\alpha\rho,\rho}} &+&
 \,b_{37}\, h_{{\mu\nu}}\, h_{{\alpha\beta}}\, h_{{\alpha\rho,\mu}}\,
h_{{\beta\rho,\nu}} &+&
 \,b_{38}\, h_{{\mu\nu}}\, h_{{\alpha\alpha}}\, h_{{\beta\mu,\rho}}\,
h_{{\beta\nu,\rho}} &+
\\
 \,b_{39}\, h_{{\mu\nu}}\, h_{{\alpha\beta}}\, h_{{\mu\rho,\nu}}\,
h_{{\alpha\beta,\rho}} &+&
 \,b_{40}\, h_{{\mu\nu}}\, h_{{\alpha\beta}}\, h_{{\alpha\rho,\mu}}\,
h_{{\beta\nu,\rho}} &+&
 \,b_{41}\, h_{{\mu\mu}}\, h_{{\alpha\nu}}\, h_{{\alpha\nu,\beta}}\,
h_{{\rho\rho,\beta}} &+
\\
 \,b_{42}\, h_{{\mu\nu}}\, h_{{\alpha\beta}}\, h_{{\alpha\beta,\mu}}\,
h_{{\rho\rho,\nu}} &+&
 \,b_{43}\, h_{{\mu\nu}}\, h_{{\alpha\mu}}\, h_{{\beta\nu,\alpha}}\,
h_{{\rho\rho,\beta}}&\left.\right)&{}&.
\end{array}
\end{equation}


{}From the gauge invariance condition, one expects that a change in the
gravitational field $\delta h_{\mu\nu}$ given by (\ref{eq3.4}), should
have no effect on the graviton-graviton tree amplitude. Imposing this
requirement and using the results mentioned in the previous equations,
we obtain a relation which can be written in correspondence with (\ref{eq2.8})
as
\begin{equation}\label{eq3.9}
\left[\delta\; tree(b_i)\right]_{1c}=-\left[\delta\; tree(\tilde a)
\right]_{1b},
\end{equation}
where $\left[\delta\; tree\right]_{1b}$ represents the gauge variation
associated with the diagram in Fig. 1b and its corresponding permutations.
It is expressed as a function of the independent coupling constants
$\tilde a_i$ left over from the analysis of the trilinear graviton vertices.
The left-hand side of eq. (\ref{eq3.9}), denotes the gauge variation
resulting from the contributions associated with the graph in Fig. 1c,
which is a function of the independent constants $b_1\cdots b_{43}$, which
parametrize the 4-graviton vertex in (\ref{eq3.8}).

We now gather together the terms with the same structure and set the
coefficients of all independent structures in (\ref{eq3.9}) equal to zero.
We then obtain a system which comprises 27 algebraically independent
equations, expressing certain linear combinations of the $b_i$ in terms of
quadratic functions of the parameters $\tilde a_i$. Clearly, this set of
independent equations cannot determine all the parameters $b_i$, nor can it
lead to any additional relations among the $\tilde a_i$.
The solution of the above system in given by a set of equations
where the 27 coefficients $b_i$ $(i=1,\, 2,\,\cdots\, ,26,\, 27)$
are expressed in terms of the remaining 16
coefficients $b_i$ $(i=28,\, 29,\,\cdots\, ,42,\, 43)$ and of the parameters
$\tilde a_i$. We write here explicitly only a few typical equations:
\begin{equation}\label{eq3.8a}
\begin{array}{lll}
8\,b_{{1}}&=-&4\,b_{{29}}-
8\,b_{{42}}+
12\,a_{{11}}a_{{8}}+
a_{{9}}a_{{13}}+
16\,a_{{16}}a_{{11}}+
6\,a_{{9}}a_{{8}}+
16\,a_{{10}}a_{{11}}+
    \\ & &
14\,a_{{9}}a_{{11}}+
5\,a_{{9}}a_{{15}}+
8\,a_{{9}}a_{{10}}+
8\,a_{{9}}a_{{16}}+
12\,a_{{15}}a_{{11}}+
6\,{a_{{9}}}^{2}+
8\,{a_{{11}}}^{2}
 \\
 8\,b_{{2}}&=-&2\,b_{{28}}-
4\,b_{{30}}-
2\,b_{{32}}-
4\,b_{{33}}+
2\,b_{{38}}-
8\,b_{{39}}-
2\,b_{{43}}-
3\,a_{{9}}a_{{8}}-
  \\ & &
3\,a_{{11}}a_{{8}}-
4\,a_{{10}}a_{{11}}-
16\,{a_{{10}}}^{2}-
4\,a_{{9}}a_{{10}}-
24\,a_{{10}}a_{{8}}+
4\,a_{{13}}a_{{11}}+
  \\ & &
4\,a_{{9}}a_{{13}}+
8\,a_{{13}}a_{{8}}-
18\,{a_{{15}}}^{2}+
10\,a_{{15}}a_{{13}}-
35\,a_{{15}}a_{{10}}-
7\,a_{{9}}a_{{15}}-
   \\ & &
7\,a_{{15}}a_{{11}}-
26\,a_{{15}}a_{{8}}+
11\,a_{{10}}a_{{13}}-
{a_{{13}}}^{2}-
4\,a_{{16}}a_{{11}}+
12\,a_{{16}}a_{{13}}-
     \\ & &
36\,a_{{15}}a_{{16}}-
32\,a_{{16}}a_{{10}}-
4\,a_{{9}}a_{{16}}-
24\,a_{{16}}a_{{8}}-
9\,{a_{{8}}}^{2}-
16\,{a_{{16}}}^{2}
 \\
\vdots &=& \vdots
 \\
2\,b_{{26}}&=&2\,b_{{28}}+
4\,b_{{33}}-
4\,b_{{38}}-
12\,a_{{9}}a_{{8}}-
4\,{a_{{9}}}^{2}-
4\,{a_{{11}}}^{2}-
12\,a_{{11}}a_{{8}}-
      \\ & &
16\,a_{{10}}a_{{11}}-
8\,a_{{9}}a_{{11}}-
16\,{a_{{10}}}^{2}-
16\,a_{{9}}a_{{10}}-
24\,a_{{10}}a_{{8}}-
8\,a_{{13}}a_{{11}}-
     \\ & &
8\,a_{{9}}a_{{13}}-
12\,a_{{13}}a_{{8}}+
4\,{a_{{15}}}^{2}-
14\,a_{{15}}a_{{13}}-
8\,a_{{15}}a_{{10}}-
4\,a_{{9}}a_{{15}}-
      \\ & &
4\,a_{{15}}a_{{11}}-
6\,a_{{15}}a_{{8}}+
16\,a_{{10}}a_{{13}}+
{a_{{13}}}^{2}-
16\,a_{{16}}a_{{11}}-
16\,a_{{16}}a_{{13}}-
       \\ & &
8\,a_{{15}}a_{{16}}-
32\,a_{{16}}a_{{10}}-
16\,a_{{9}}a_{{16}}-
24\,a_{{16}}a_{{8}}-
9\,{a_{{8}}}^{2}-
16\,{a_{{16}}}^{2}
 \\
16\,b_{{27}}&=&
8\,b_{{42}}-
4\,a_{{9}}a_{{16}}-
4\,{a_{{9}}}^{2}-
3\,a_{{9}}a_{{15}}-
4\,a_{{9}}a_{{10}}-
3\,a_{{9}}a_{{8}}-
12\,a_{{9}}a_{{11}}-
                 \\ & &
12\,a_{{11}}a_{{8}}-
16\,a_{{16}}a_{{11}}-
16\,a_{{10}}a_{{11}}-
12\,a_{{15}}a_{{11}}-
8\,{a_{{11}}}^{2}.
\end{array}
\end{equation}

Although much more complicated in detail, these relations are basically
similar to the one encountered in the Yang-Mills case
[see eq. (\ref{eq2.11})].
The crucial difference occurs when attempting, in
parallel to the procedure used in the Yang-Mills case, to apply the gauge
invariance condition to the 5-graviton tree amplitude. Now, there exists
a basic 5-graviton vertex, shown in Fig 2c, which must be parametrized
in terms of the most general sum of independent products involving five
graviton fields with two derivatives. This parametrization can be done in terms
of a very large number of new constants, which we denote by the set
$c_i$. Following closely the analysis after equation (\ref{eq3.9}), is is clear
that the Abelian gauge invariance condition will merely lead to some
relations expressing certain $c_i$ in terms of the remaining $c_i$ and of
the parameters $\tilde a_i$ and $b_i$ left over from the previous analysis.

It is evident that this behavior is quite general, in view of the fact that
the graviton self-couplings occur to all orders. We thus conclude that the
constraint of Abelian gauge invariance of the physical tree amplitudes does
not determine completely the from of the nonlinear graviton interactions.
It is only when we impose the condition that the theory should be invariant
under (infinitesimal) non-Abelian gauge transformations
\begin{equation}\label{eq3.10}
h_{\mu\nu}\rightarrow
h_{\mu\nu} + \xi_{\mu,\nu} + \xi_{\nu,\mu} +
\kappa \left(
\xi^{\sigma}_{,\mu} h_{\sigma\nu}+\xi^{\sigma}_{,\nu} h_{\sigma\mu}+
\xi^{\sigma} h_{\mu\nu,\sigma}
\right).
\end{equation}
that it becomes essentially determined.
For example, using the parametrization given by (\ref{eq3.5}), we find
in this case for the trilinear graviton vertex:
\begin{equation}\label{eq3.11}
\begin{array}{lllll}
a_1 = 1+a_{14} \; ,\;\; &  2\,a_2 =-{3}-2\,a_{15}\; ,\;\;&{}
a_3 =1-a_{14} \; ,\;\; &  a_4 =a_3 \; ,\;\;&{}\\
2\,a_5 =-{1}\; ,\;\; &  a_6 =-1  \; ,\;\;&{}
4\, a_7 = 1\; ,\;\; &
2\,a_8=-{1}-2\,a_{15} & {}\\
a_9=-1 \; ,\;\; &  2\, a_{10}={1} \; ,\;\;&{}
a_{11}=1 \; ,\;\; &  2\, a_{12}={1} \; ,\;\;&{}\\
2\, a_{13}=2\, a_{15}-{1} \; ,\;\; &
4\, a_{16}=-{1} \; \;\;&{}.
\end{array}
\end{equation}
We remark that the structures which multiply $a_{14}$ and $a_{15}$ add up to
total derivative terms.
Since total derivatives are not relevant for our purpose, this
result is equivalent to the one obtained from the Einstein's general
relativity. Then, the theory becomes consistent with the existence of a locally
conserved energy-momentum ``tensor'' of matter
{and~gravitation\cite{weinberg}}.

\acknowledgements{We would like to thank CNPq (Brasil) for a grant. J. F. is
grateful to Prof. J. C. Taylor for a very helpful correspondence} and for
reading the manuscript.

%
%
%
%
%

\end{document}